\documentclass[twocolumn,showpacs]{revtex4}
\usepackage{amsmath,amsthm,amssymb,amscd,float}
\usepackage{graphicx}
\newcommand{\al}{\alpha}
\newcommand{\be}{\beta}
\newcommand{\de}{\delta}
\newcommand{\ep}{\epsilon}

\newcommand{\ga}{\gamma}

\newcommand{\si}{\sigma}

\newcommand{\De}{\Delta}

\newcommand{\La}{\Lambda}
\newcommand{\Si}{\Sigma}
\newcommand{\bk}{\mathbf{k}}

\newcommand{\bm}{\mathbf{m}}
\newcommand{\bn}{\mathbf{n}}

\newcommand{\bs}{\mathbf{s}}
\newcommand{\bx}{\mathbf{x}}
\newcommand{\bV}{\mathbf{V}}
\newcommand{\bnu}{{\boldsymbol{\nu}}}

\newcommand{\bxi}{{\boldsymbol{\xi}}}

\newcommand{\tK}{\widetilde{K}}

\newcommand{\tS}{\tilde{S}}

\newcommand{\tih}{\tilde{h}}

\newcommand{\NN}{{\mathbb N}}
\newcommand{\RR}{{\mathbb R}}
\newcommand{\CC}{{\mathbb C}}

\newcommand{\cE}{{\mathcal E}}

\newcommand{\cH}{{\mathcal H}}

\newcommand{\cP}{{\mathcal P}}

\newcommand{\cZ}{{\mathcal Z}}
\def\Esc{E^{\mathrm{sc}}}
\def\Hsc{H_{\mathrm{sc}}}
\def\Zsc{Z_{\mathrm{sc}}}
\newcommand{\pa}{\partial}

\newcommand{\abs}[1]{\left|#1\right|}
\def\ket#1{|#1\rangle}

\let\ni\noindent
\newcommand{\ms}{\mspace{1mu}}
\renewcommand{\le}{\leqslant}

\renewcommand{\geq}{\geqslant}
\newcommand{\qbinom}[3]{\genfrac{[}{]}{0pt}{}{\,#1\,}{#2}_{#3}}
\newcommand{\erf}{\operatorname{erf}}

\newcommand{\tr}{\operatorname{tr}}

\newcommand{\e}{\mathrm{e}}

\newcommand{\diff}{\mathrm{d}}
\newcommand{\smax}{s_{\mathrm{max}}}
\newcounter{ex}
\def\cond{\stepcounter{ex}\hskip-.75cm
\makebox[.55cm][r]{(\roman{ex})}\hskip.2cm}
\begin{document}
\title{Exactly solvable $D_N$-type quantum spin models with long-range interaction}
\author{B. \surname{Basu-Mallick}}%
\email{bireswar.basumallick@saha.ac.in}
\affiliation{Theory Group, Saha Institute of Nuclear Physics, 1/AF Bidhan
  Nagar, Kolkata 700 064, India}
\author{F. \surname{Finkel}}%
\email{ffinkel@fis.ucm.es}
\author{A. \surname{Gonz\'alez-L\'opez}}
\email[Corresponding author. Electronic address: ]{artemio@fis.ucm.es}
\affiliation{Departamento de F\'{\i}sica
  Te\'{o}rica II, Universidad Complutense, 28040 Madrid, Spain}
\date{September 23, 2008}
\begin{abstract}
  We derive the spectra of the $D_N$-type Calogero (rational) su($m$) spin model, including the
  degeneracy factors of all energy levels. By taking the strong coupling limit of this model, in
  which its spin and dynamical degrees of freedom decouple, we compute the exact partition
  function of the su($m$) Polychronakos--Frahm spin chain of $D_N$ type. With the help of this
  partition function we study several statistical properties of the chain's spectrum, such as the
  density of energy levels and the distribution of spacings between consecutive levels.
\end{abstract}
\maketitle
%
%
\section{Introduction}

Recent studies of quantum integrable dynamical models and spin chains with long-range
interactions~\cite{Ca71,Su71,Su72, OP83, Ha88,Sh88,Po93,Ha96} have not only enriched our
understanding of strongly correlated many-particle systems in one dimension, but also influenced
several branches of mathematics in a significant way. In particular, it is found that this class
of quantum integrable systems have close connections with apparently diverse subjects like
generalized exclusion statistics~\cite{Ha96,MS94,Po06}, quantum Hall effect~\cite{AI94}, quantum
electric transport in mesoscopic systems~\cite{BR94,Ca95}, random matrix theory~\cite{TSA95},
multivariate orthogonal polynomials~\cite{Fo94,UW97,BF97npb} and Yangian quantum
groups~\cite{BGHP93,Hi95npb,BK98}. The interest in quantum integrable models with long-range
interaction was initiated by a seminal work of Calogero~\cite{Ca71}, where the exact spectrum of
an $N$-particle system on a line with two-body interactions inversely proportional to the square
of their distances and subject to a confining harmonic potential was computed in closed form. An
exactly solvable trigonometric variant of the rational model introduced by Calogero was proposed
shortly afterwards by Sutherland~\cite{Su71,Su72}. The particles in this so-called Sutherland
model move on a circle, with two-body interactions proportional to the inverse square of their
chord distances. Subsequently, Olshanetsky and Perelomov established the existence of an
underlying $A_N$ root system structure for both the Calogero and Sutherland models, and
constructed generalizations thereof associated with other classical (extended) root systems like
$B_N$, $C_N$ and $BC_N$~\cite{OP83}.

In a parallel development, Haldane and Shastry found an exactly solvable quantum
spin-$\frac{1}{2}$ chain with long-range interactions, whose ground state coincides with the
$U\rightarrow \infty $ limit of Gutzwiller's variational wave function for the Hubbard model, and
provides a one-dimensional realization of the resonating valence bond state~\cite{Ha88,Sh88}. The
lattice sites of this su$(2)$ Haldane--Shastry (HS) spin chain are equally spaced on a circle, all
spins interacting with each other through pairwise exchange interactions inversely proportional to
the square of their chord distances. A close relation between the HS chain and the su($m$) spin
generalization of the original ($A_N$-type) Sutherland model~\cite{HH92,HW93,MP93}, which leads to
many quantitative predictions, was subsequently established through the so-called ``freezing
trick''~\cite{Po93,SS93}. More precisely, it is found that in the strong coupling limit the
particles in the spin Sutherland model ``freeze'' at the coordinates of the equilibrium position
of the scalar part of the potential, and the dynamical and spin degrees of freedom decouple. The
equilibrium coordinates coincide with the equally spaced lattice points of the HS spin chain, so
that the decoupled spin degrees of freedom are governed by the Hamiltonian of the su($m$) HS
model. Moreover, in this freezing limit the conserved quantities of the spin Sutherland model
immediately yield those of the HS spin chain, thereby explaining its complete integrability. By
applying the freezing trick to the $A_N$-type rational Calogero model with spin degrees of
freedom, a new integrable spin chain with long-range interaction was constructed in
Ref.~\cite{Po93}. The sites of this chain ---commonly known in the literature as the Polychronakos
or Polychronakos--Frahm (PF) spin chain--- are unequally spaced on a line, and in fact coincide
with the zeros of the $N$-th Hermite polynomial~\cite{Fr93}.

The powerful technique of the freezing trick was subsequently used to compute the exact partition
functions of both the su($m$) PF spin chain~\cite{Po94} and the su($m$) HS chain~\cite{FG05}, the
$BC_N$ counterparts of these chains~\cite{BFGR08,EFGR05}, and their supersymmetric
extensions~\cite{BUW99,BB06,BFGR08npb}. The exact computation of the partition functions of these
quantum integrable spin chains has opened up the exciting possibility of studying various
statistical properties of their energy spectra. Indeed, it is found that for a large number of
lattice sites the energy level density of such chains follows the Gaussian distribution with a
high degree of accuracy~\cite{FG05,BFGR08,EFGR05,BFGR08epl,BB06,BFGR08npb}. It has also been
observed that the distribution of the (normalized) spacings between consecutive energy levels of
these chains is not of Poisson type, as might be expected in view of a well-known conjecture of
Berry and Tabor~\cite{BT77}. An analytical expression, which explains the unexpected distribution
of spacings between consecutive energy levels in the above mentioned chains, has recently been
derived using only a few simple properties of their spectra~\cite{BFGR08}.

Our aim in this article is first of all to analyze the spectra of the su($m$) spin Calogero model
of $D_N$ type. We shall then apply the freezing trick to compute the exact partition function of
the $D_N$ version of the PF spin chain, and use this partition function to study various
statistical properties of the chain's spectrum. It should be stressed that, although the
Hamiltonian of the $D_N$-type su($m$) spin Calogero model can be obtained by setting to zero one
of the coupling constants of their $BC_N$ counterparts, this fact does not allow one to find out
all physically relevant properties of the $D_N$ model as a limiting case of its $BC_N$ version.
For example, as will be explained in Section~\ref{themodel}, the configuration space of the
$D_N$-type spin Calogero model differs quite significantly from its $BC_N$ counterpart. A more
drastic change occurs in the Hilbert space of the $D_N$ model, which gets ``doubled'' in
comparison with the $BC_N$ one. More precisely, the Hilbert space of the $D_N$ spin Calogero model
can be expressed as a direct sum of the Hilbert spaces associated to two different $BC_N$ models
with opposite ``chiralities''. These remarkable properties of the $D_N$ model indicate that it is
a ``singular limit'' of the corresponding $BC_N$ model, worthy of consideration in its own right.

The paper is organized as follows. In Section~\ref{themodel} we introduce the su($m$) spin
Calogero model of $D_N$ type and construct its associated (antiferromagnetic) spin chain by means
of the freezing trick, discussing their relation with their $BC_N$ counterparts.
Section~\ref{specPF} is devoted to the evaluation of the spectrum of the spin Calogero model of
$D_N$ type, which is then used to compute in closed form the partition function of its associated
spin chain applying the freezing trick. We also show how to express this partition function in
terms of those of the PF chains of type $A$ and $B$. In Section~\ref{stat} we make use of the
closed-form expressions for the partition function of the PF chain of $D_N$ type to analyze
several statistical properties of its spectrum. We show that ---as is the case with other chains
of HS type--- the level density follows with great accuracy the Gaussian law when the number of
lattice sites is sufficiently large. We also prove that the cumulative distribution of spacings
between consecutive levels follows the same ``square root of a logarithm'' law obeyed by the PF
chain of types A and B and by the original HS chain. This provides further confirmation of the
fact that spin chains of HS type are exceptional integrable systems from the point of view of the
Berry--Tabor conjecture. Finally, in Section~\ref{ferro} we outline the generalization of the
above results to the ferromagnetic chain and its associated spin dynamical model.

\section{The model}\label{themodel}
Since the su($m$) spin Calogero model of $D_N$ type is closely related to its $BC_N$ counterpart,
we shall start by briefly reviewing the latter model, whose Hamiltonian is given by~\cite{BFGR08}
\begin{align}
H^{(\mathrm{B})}=&-\sum_i\pa_{x_i}^2
+a\sum_{i\neq j}\bigg[
\frac{a+S_{ij}}{(x_{ij}^-)^2}+\frac{a+\tS_{ij}}{(x_{ij}^+)^2}\bigg]
\notag\\ &{}+b \sum_i \frac{b -\epsilon S_i}{x_i^2} +
\frac{a^2}4\,r^2 \,. \label{BH}
\end{align}
Here the sums run from $1$ to $N$ (as always hereafter, unless otherwise stated), $a>1/2$, $b>0$,
$\epsilon =\pm 1$, $x_{ij}^{\pm}=x_i\pm x_j$, $r^2=\sum_ix_i^2$, $S_{ij}$ is the operator which permutes
the $i$-th and $j$-th spins, $S_i$ is the operator reversing the $i$-th spin, and
$\tS_{ij}=S_iS_jS_{ij}$. Note that the spin operators $S_{ij}$ and $S_i$ can be expressed in terms
of the fundamental su($m$) spin generators $J^\al_k$ at the site $k$ (with the normalization
$\tr(J^\al_kJ^{\ga}_k)=\frac12\de^{\al\ga}$) as
\[
S_{ij}=\frac1m+2\sum_{\al=1}^{m^2-1}J_i^\al J_j^\al\,,\qquad S_i=\sqrt{2m}\,J^1_i\,.
\]
The configuration space of the Hamiltonian~\eqref{BH} can be taken as one of the Weyl chambers of
the $BC_N$ root system, i.e., one of the maximal open subsets of $\RR^N$ on which the linear
functionals $x_i\pm x_j$ and $x_i$ have constant signs. We shall choose as configuration space the
principal Weyl chamber
\begin{equation}
  C^{(\mathrm{B})}=\{\bx\equiv(x_1,\dots,x_N): 0<x_1<x_2<\cdots<x_N\}\,.
  \label{BC}
\end{equation}
The spectrum of the $BC_N$ spin Calogero model, including the degeneracy factors of all energy
levels, has been determined by constructing a (non-orthonormal) basis of the Hilbert space in
which the Hamiltonian~\eqref{BH} is triangular~\cite{BFGR08}. By setting $b=\beta a$ and taking
the limit $a\rightarrow \infty$ in the Hamiltonian~\eqref{BH}, one can obtain the su($m$) PF spin
chain of $BC_N$ type, with Hamiltonian given by
\begin{equation}\label{BcH}
{\cH}^{(\mathrm{B})}=\sum_{i\neq j}\bigg[
\frac{1+S_{ij}}{(\xi_i-\xi_j)^2}+\frac{1+\tS_{ij}}{(\xi_i+\xi_j)^2}\bigg]\,
+ \beta \sum_i\frac{1-\epsilon S_i}{\xi_i^2} \,.
\end{equation}
Here $\beta $ is a positive real parameter, and the lattice sites $\xi_i$ can be expressed in
terms of the zeros $y_i$ of the Laguerre polynomial $L_{N}^{\beta -1} $ as $y_i=\xi_{i}^2/2$. The
exact partition function of the spin model~\eqref{BcH} has also been recently computed with the
help of freezing trick~\cite{BFGR08}.

The Hamiltonian of the su($m$) spin Calogero model of $D_N$ type is
obtained by setting $b=0$ in its $BC_N$ counterpart~\eqref{BH}, namely
\begin{equation}
  H=-\sum_i\pa_{x_i}^2 + \frac{a^2}4\,r^2 \,
  +a\sum_{i\neq j}\bigg[
  \frac{a+S_{ij}}{(x_{ij}^-)^2}+\frac{a+\tS_{ij}}{(x_{ij}^+)^2}\bigg]
  \,. \label{H}
\end{equation}
As configuration space of the Hamiltonian~\eqref{H} we can take again one of the Weyl chambers of
the $D_N$ root system. For instance, the choice $x_1<\cdots <x_N$ determines all the differences
$x_i-x_j$. If we also require that $x_1+x_2>0$ the sign of all the sums $x_i+x_j$ is determined as
well. Indeed, $|x_1|<x_2$ implies that $|x_1|<x_j$ for all $j=2,\dots, N$, so that $x_1+x_j>0$ for
$j>1$, while the sums $x_i+x_j$ with $i,j>2$ and $i\ne j$ are clearly positive on account of the
positivity of $x_k$ with $k>1$. Thus we can take as configuration space of $H$ the open set
\begin{equation}
  C=\{\bx\equiv(x_1,\dots,x_N): \abs{x_1}<x_2<\cdots<x_N\}\,,
  \label{C}
\end{equation}
which is just the principal Weyl chamber of the $D_N$ root system. It is interesting to observe
that this configuration space contains its $BC_N$ counterpart~\eqref{BC} as a subset.

The Hamiltonian of the su($m$) PF chain of $D_N$ type can be obtained from the spin
Hamiltonian~\eqref{H} in the limit $a\to\infty$ by means of the freezing trick. More precisely,
since
\[
H=-\sum_i\pa_{x_i}^2+a^2 U+O(a)\,,
\]
with
\begin{equation}\label{U}
U(\bx)=\sum_{i\neq j}\bigg[\frac1{(x_{ij}^{-})^2}+\frac1{(x_{ij}^{+})^2}\bigg]
+\frac{r^2}4\,,
\end{equation}
when the coupling constant $a$ tends to infinity the particles in the spin dynamical
model~\eqref{H} concentrate at the coordinates $\xi_i$ of the minimum $\bxi$ of the potential $U$
in $C$. From the identity
\begin{equation}
  \label{Hepchain}
  H=\Hsc+a\,\widetilde\cH(\bx)\,,
\end{equation}
where
\begin{equation}\label{Hsc}
\Hsc=-\sum_i\pa_{x_i}^2+\frac{a^2}4\,r^2+a(a-1)\sum_{i\neq j}\bigg[
\frac1{(x_{ij}^-)^2}+\frac1{(x_{ij}^+)^2}\bigg]
\end{equation}
and
\[
\widetilde\cH =\sum_{i\ne
  j}\bigg[\frac{1+S_{ij}}{(x_i-x_j)^2}+\frac{1+\tS_{ij}}{(x_i+x_j)^2}\bigg]\,,
\]
it follows that in the limit $a\to\infty$ the internal degrees of freedom of $H$ are governed by
the Hamiltonian $\cH=\widetilde\cH(\bxi)$, explicitly given by
\begin{equation}\label{cH}
\cH=\sum_{i\neq j}\bigg[
\frac{1+S_{ij}}{(\xi_i-\xi_j)^2}+\frac{1+\tS_{ij}}{(\xi_i+\xi_j)^2}\bigg]\,.
\end{equation}
Equation~\eqref{cH} is the Hamiltonian of the (antiferromagnetic) su($m$) PF chain of $D_N$ type,
whose sites $\xi_i$ are the coordinates of the unique minimum $\bxi$ of the scalar
potential~\eqref{U} in the open set~\eqref{C}. The existence of this minimum follows from the fact
that $U$ tends to $+\infty$ on the boundary of $C$ and as $r\to\infty$, and its uniqueness was
established in~Ref.~\cite{CS02} by expressing the potential $U$ in terms of the logarithm of the
ground state $\rho$ of the scalar $D_N$ Calogero model $\Hsc$, given by
\begin{equation}\label{rho}
\rho(\bx)=\e^{-\frac a4\ms r^2}\prod_{i<j}{\big|x_i^2-x_j^2\big|}^a.
\end{equation}
As shown in the latter reference, the sites $\xi_i$ coincide with the coordinates of the (unique)
critical point of $\log\rho$ in $C$, and therefore satisfy the nonlinear system
\[
\xi_i\bigg(
\sum_{j;j\ne i}\frac{1}{\xi_i^2-\xi_j^2}-\frac14
\bigg)=0\,,\qquad i=1,\dots,N\,.
\]
The numbers $\xi_i$ cannot be all different from zero, since in that case we would obtain the
contradiction
\[
0 = \sum_{j\ne i}\frac{1}{\xi_i^2-\xi_j^2}-\frac N4 = -\frac N4\,.
\]
Hence $\xi_i=0$ for some $i$, and since $(\xi_1,\dots,\xi_N)$ lies in $C$
we must have
\begin{subequations}
  \label{xis}
  \begin{equation}
  \label{xi1}
  \xi_1=0\,,
\end{equation}
while the remaining $N-1$ sites should satisfy the condition
\begin{equation}
  \label{xi2}
  \sum_{j;j\ne i}\frac{1}{\xi_i^2-\xi_j^2}=\frac14\,,
  \qquad i=2,\dots,N\,.
\end{equation}
\end{subequations}
Substituting Eq.~\eqref{xi1} into \eqref{xi2} one obtains
\begin{equation}
  \label{xisfinal}
  \sum_{\substack{j=2\\j\ne i}}^N\frac{1}{\xi_i^2-\xi_j^2}=\frac14-\frac1{\xi_i^2}\,,
  \qquad i=2,\dots,N\,.
\end{equation}

It is interesting to compare the above condition satisfied by the
nonzero $\xi_i$'s with the relation
\begin{equation}
  \label{zl}
  \sum_{\substack{j=1\\j\ne i}}^{M}
  \frac{1}{(y_i-y_j)}=\frac12-\frac{\beta}{2y_i}\,,
\end{equation}
obeyed by the zeros $y_i$ of the Laguerre polynomial $L_{M}^{\beta -1} $~\cite{Ah78}. It is
evident that Eq.~\eqref{xisfinal} reduces to Eq.~\eqref{zl} when $M=N-1$, $\be=2$ and
$y_{i}=\xi_{i-1}^2/2$. We therefore conclude that the sites $\xi_2<\dots<\xi_N$ are expressed in
terms of the $N-1$ zeros $y_1<\cdots<y_{N-1}$ of the Laguerre polynomial $L_{N-1}^1$ by
$\xi_i=\sqrt{2y_{i-1}}$. On the other hand, it has already been mentioned that the lattice sites
of the PF model of $BC_N$ type~\eqref{BcH} are expressed in terms of the zeros of the Laguerre
polynomial $L_{N}^{\beta -1} $ by $y_i=\xi_{i}^2/2$. Since the potential $U$ in Eq.~\eqref{U} is
obtained from its $BC_N$ counterpart in the limit $\beta \rightarrow 0$, we could also have argued
that the lattice site $\xi_i$ of the $D_N$-type PF model is the square root of twice the $i$-th
zero of $L_N^{-1}$ for $i=1,\dots,N$. The equivalence of both characterizations is substantiated
by the well-known identity $L_N^{-1}(y)=-y\,L_{N-1}^1(y)/N$, cf.~\cite{CS02}.

It is worth pointing out that, even though the lattice sites of the $BC_N$-type PF chain coincide
with their $D_N$ counterparts in the limit $\beta \rightarrow 0$, the Hamiltonian~\eqref{BcH} of
the PF chain of $BC_N$ type does not reduce to its $D_N$ variant~\eqref{cH} in the same limit. To
establish this fact, note first that all roots of the equation $L_{N}^{\beta -1}(y)=0$ except the
smallest one tend to a finite nonzero value in the limit $\beta \rightarrow 0$. As a result, terms
like $\beta (1-\epsilon S_i)/\xi_i^2$, which appear in the r.h.s.\ of Eq.~\eqref{BcH}, vanish for
$i=2,\dots,N$. We next examine the behavior of the smallest root $\xi_1$ of the equation
$L_{N}^{\beta -1}(y)=0$. It can be shown~\cite{BFGR08} that the zeros of $L_{N}^{\beta -1} $
satisfies the relation
\begin{equation}
  \label{zl1}
  \beta \sum_{j=2}^{N}
  \frac{1}{y_j}=N-\frac{\beta}{y_1}\,.
\end{equation}
Since the l.h.s.\ of this equation vanishes in the limit $\beta\to0$, the r.h.s.\ yields
$\lim_{\beta \rightarrow 0}(2\beta/\xi_1^2 )= N$. Substituting this limiting value in
Eq.~\eqref{BcH} we find that the Hamiltonians of the $BC_N$- and $D_N$-type PF spin chains are
related by
\begin{equation}
  \label{rela}
 \lim_{\beta \rightarrow 0} {\cH}^{(\mathrm{B})} = {\cH} + \frac{N}{2}
 (1-\epsilon S_1) \,.
\end{equation}
It is interesting to observe that the second term in the r.h.s.\ of the previous equation may be
interpreted as an ``impurity'' interaction at the left end of the $BC_N$ spin chain.

\section{Spectrum and partition function}\label{specPF}

We shall start by deriving the spectra and partition functions of the $D_N$-type su($m$) spin
Calogero model~\eqref{H} and its scalar counterpart~\eqref{Hsc}. Since the spin and dynamical
degrees of freedom of the Hamiltonian~\eqref{H} decouple in the freezing limit $a\to\infty$, by
Eq.~\eqref{Hepchain} its eigenvalues are approximately given by
\begin{equation}\label{EEE}
E_{ij}\simeq E^{\mathrm{sc}}_i+a\cE_j\,,\qquad a\gg1\,,
\end{equation}
where $E^{\mathrm{sc}}_i$ and $\cE_j$ are two arbitrary eigenvalues of $\Hsc$ and $\cH$,
respectively. The asymptotic relation~\eqref{EEE} immediately yields the following exact formula
for the partition function $\cZ$ of the $D_N$-type PF spin chain~\eqref{cH}:
\begin{equation}\label{ZZZ}
\cZ(T)=\lim_{a\to\infty}\frac{Z(aT)}{\Zsc(aT)}\,,
\end{equation}
where $Z$ and $\Zsc$ are the partition functions of $H$ and $\Hsc$, respectively. Inserting
the expressions for the partition functions $Z$ and $\Zsc$ in the latter equation
we shall obtain an explicit formula for the partition function $\cZ$ of the
chain~\eqref{cH}.

In order to determine the spectra of the corresponding Hamiltonians $H$ and $\Hsc$ in
Eqs.~\eqref{H} and \eqref{Hsc}, following Ref.~\cite{BFGR08} we introduce the auxiliary operator
\begin{multline}
  H'=-\sum_i\pa_{x_i}^2+\frac{a^2}4\,r^2\\
  +\sum_{i\neq j}\bigg[ \frac a{(x_{ij}^-)^2}(a-K_{ij})+\frac
  a{(x_{ij}^+)^2}(a-\tK_{ij})\bigg]\,,\label{Hp}
\end{multline}
where $K_{ij}$ and $K_i$ are coordinate permutation and sign reversing operators, defined by
\begin{align*}
&(K_{ij}f)(x_1,\dots,x_i,\dots,x_j,\dots,x_N)\\
&\hspace{8.5em}=f(x_1,\dots,x_j,\dots,x_i,\dots,x_N)\,,\\
&(K_i f)(x_1,\dots,x_i,\dots,x_N)=f(x_1,\dots,-x_i,\dots,x_N)\,,
\end{align*}
and $\tK_{ij}=K_iK_jK_{ij}$. We then have the obvious relations
\begin{subequations}
  \label{HpHs}
\begin{align}
H&=H'\big|_{K_{ij}\to-S_{ij},K_i\to\ep S_i}\,,\label{HepHp}\\
\Hsc&=H'\big|_{K_{ij}\to1,K_i\to\ep}\,,\label{HscHp}
\end{align}
\end{subequations}
where $\ep$ can take \emph{both} values $\pm1$. On the other hand, the spectrum of $H'$ is easily
computed by noting that this operator can be written in terms of the rational Dunkl operators of
$D_N$ type~\cite{Du98}
\begin{equation}
J_i^-=\pa_{x_i}+a\sum_{j\neq i}\bigg[\frac1{x_{ij}^-}\,(1-K_{ij})
+\frac1{x_{ij}^+}\,(1-\tK_{ij})\bigg]
\,,\label{J-}
\end{equation}
$i=1,\dots,N$, as follows~\cite{FGGRZ01b}:
\begin{equation}\label{Hp2}
H'=\rho\Big[-\sum_i\big(J_i^-\big)^2+a\sum_i x_i\pa_{x_i}+E_0\Big]\rho^{-1}\,,
\end{equation}
where
\begin{equation}\label{E0}
E_0=Na\Big(a(N-1)+\frac12\Big)\,.
\end{equation}
Since the Dunkl operators~\eqref{J-} map any monomial $\prod_i x_i^{n_i}$ into a polynomial of
total degree $n_1+\cdots+n_N-1$, by Eq.~\eqref{Hp2} the operator $H'$ is represented by an upper
triangular matrix in the (non-orthonormal) basis with elements
\begin{equation}
\phi_\bn=\rho\prod_ix_i^{n_i},\:
\quad \bn\equiv(n_1,\dots,n_N)\in{(\NN\cup\{0\})}^N,
\end{equation}
ordered according to the total degree
$\vert\bn\vert\equiv n_1+\cdots+n_N$ of the
monomial part. More precisely,
\begin{equation}
  \label{Hpphin}
H'\phi_\bn=
E'_\bn\phi_\bn+\sum_{\vert\bm\vert<\vert\bn\vert}c_{\bm\bn}\phi_\bm\,,
\end{equation}
where
\begin{equation}\label{Ep}
E'_\bn=a\ms\vert\bn\vert+E_0
\end{equation}
and the coefficients $c_{\bm\bn}$ are real constants.

We shall now construct a basis of the Hilbert space of the Hamiltonian $H$ in which this operator
is also represented by an upper triangular matrix. To this end, let us denote by
$\Si\approx(\CC^m)^{\otimes N}$ the Hilbert space of the su($m$) internal degrees of freedom, and
let
\[
\ket\bs\equiv\ket{s_1,\dots,s_N},\quad s_i=-M,-M+1,\dots,M\equiv{\textstyle\frac{m-1}2},
\]
be an arbitrary element of the canonical (orthonormal) basis in this space. Due to the
impenetrable nature of the singularities of the Hamiltonian~\eqref{H}, its Hilbert space is the
set $L_0^2(C)\otimes\Si$ of spin wave functions square integrable on the open set $C$ which vanish
sufficiently fast on the singular hyperplanes $x_i\pm x_j=0$, $1\le i<j\le N$. It can be shown,
however, that $H$ is equivalent to its natural extension to the subspace of
$L_0^2(\RR^N)\otimes\Si$ consisting of spin wave functions antisymmetric under particle
permutations and symmetric under sign reversals of an \emph{even} number of coordinates and spins.
(This is essentially due to the fact that any point in $\RR^N$ not lying on the singular subset
$x_i\pm x_j=0$, $1\le i<j\le N$, can be mapped in a unique way to a point in $C$ via a suitable
element of the $D_N$ Weyl group, which is generated by coordinate permutations and sign reversals
of an even number of coordinates~\cite{Hu72}.) We can therefore assume without loss of
generality~that the Hilbert space of $H$ is the closure of the subspace spanned by the functions
\begin{equation}\label{psis}
  \psi_{\bn,\bs}^\ep(\bx)=\La^\ep\big(\phi_\bn(\bx)\ket\bs\big)\,,
  \qquad\ep=\pm1\,,
\end{equation}
where $\La^\ep$ denotes the projector on states antisymmetric under simultaneous permutations of
spatial and spin coordinates, and with parity $\ep$ under sign reversals of coordinates and spins.
The latter functions are linearly independent, and hence form a (non-orthonormal) basis of the
Hilbert space of $H$, provided that the quantum numbers $\bn$ and $\bs$ satisfy the following
conditions:\smallskip

{\leftskip.75cm\parindent=0pt%
\cond $n_1\geq\cdots\geq n_N$.

\cond $s_i>s_j$ whenever $n_i=n_j$ and $i<j$.

\cond $s_i\geq 0$ for all $i$, and $s_i>0$ if $(-1)^{n_i}=-\ep$.\smallskip

} \ni The first two conditions are a consequence of the antisymmetry of the states~\eqref{psis}
under particle permutations, while the last condition is due to the fact that $\psi_{\bn,\bs}^\ep$
must have parity $\ep$ under sign reversals. It should be noted that the Hilbert space $\mathbf V$
of the Hamiltonian $H$ just defined can be written as the direct sum
\begin{equation}\label{sum}
  \mathbf{V} = \mathbf{V}_{+} \oplus \mathbf{V}_{-} \,,
\end{equation}
where the subspace $\mathbf{V}_{\ep}$ is the closure of the span of the basis vectors
$\psi_{\bn,\bs}^\ep(\bx)$. Within each subspace $\bV_\ep$, a partial ordering among these basis
vectors may again be defined by the degree $\vert\bn\vert$. We shall now show that the Hamiltonian
$H$ is represented by an upper triangular matrix in this basis (and thus by a direct sum of two
upper triangular matrices in the total Hilbert space $\mathbf{V}$). Indeed, since
$K_{ij}\La^\ep=-S_{ij}\La^\ep$ and $K_i\La^\ep=\ep S_i\La^\ep$, it follows that
$H\La^\ep=H'\La^\ep$. Using this identity, Eq.~\eqref{Hpphin}, and the fact that $H'$ obviously
commutes with $\La^\ep$, we have
\begin{align}
H\psi^\ep_{\bn,\bs}&=H'\psi^\ep_{\bn,\bs}
=\La^\ep\big((H'\phi_\bn)\ket\bs\big)\notag\\
&=E'_\bn\psi^\ep_{\bn,\bs}+
\sum_{\vert\bm\vert<\vert\bn\vert}c_{\bm\bn}\psi^\ep_{\bm,\bs}\,.
\label{tri}
\end{align}
Suppose now that both $\bn$ and $\bs$ satisfy conditions (i)--(iii) above, so that
$\psi^\ep_{\bn,\bs}$ belongs to the basis of $\bV_\ep$ under consideration.
Although a given pair of quantum numbers $(\bm$, $\bs)$ in the r.h.s.~of the previous equation
need not satisfy these conditions, it is easy to see that
there is a permutation $\pi_\bm$ such that
$\bm'\equiv\pi_\bm(\bm)$ and $\bs'\equiv\pi_\bm(\bs)$
do satisfy (i)--(iii). Since $\psi_{\bm,\bs}^\ep$ differs from the basis vector $\psi_{\bm',\bs'}^\ep$
at most by a sign, and $\vert\bm'\vert=\vert\bm\vert<\vert\bn\vert$, our claim
follows directly from Eq.~\eqref{tri}. Moreover, the latter equation and Eq.~\eqref{Ep}
imply that the eigenvalues of the spin Calogero Hamiltonian~\eqref{H} are
given by
\begin{equation}\label{Ens}
E^\ep_{\bn,\bs}=a\ms\vert\bn\vert+E_0\,,
\end{equation}
where $\ep=\pm 1$ and $\bn,\bs$ satisfy conditions (i)--(iii) above. Since the numerical value of
$E^\ep_{\bn,\bs}$ is independent of $\bs$ and $\ep$, the energy associated with a quantum number
$\bn$ will be highly degenerate in general. For any given $\bn$, this degeneracy factor $d_\bn$
can be found by counting the numbers $d_{\bn}^\ep$ of independent spin states $\ket\bs$ satisfying
conditions (ii) and (iii) for each case $\ep=+1$ and $\ep=-1$, and finally taking the sum of these
two numbers. Explicit expressions for such degeneracy factors will be given shortly when computing
the partition function of the model.

It is important at this point to elucidate the connection between the Hilbert spaces of the
$D_N$-type spin Calogero model and its $BC_N$ counterpart. The key fact in this respect is that
the $D_N$ Hamiltonian~\eqref{H} does not depend on the discrete parameter $\ep$. Consequently, as
shown in Eq.~\eqref{psis}, we can use both projectors $\La^{+}$ and $\La^{-}$ for constructing
the Hilbert space. On the other hand, since $\ep$ appears explicitly in the Hamiltonian of the
$BC_N$ spin Calogero model~\eqref{BH}, for any given value of $\ep$ only the corresponding
projector $\La^\ep$ can be used to construct the Hilbert space~\cite{BFGR08}. Moreover, when $b=0$
this Hilbert space is essentially the subspace $\mathbf{V}_{\ep}$ of $\mathbf{V}$ in
Eq.~\eqref{sum}. Thus the presence of $\ep$ in the Hamiltonian of the $BC_N$ spin Calogero model
effectively introduces a ``chirality'' in this system. By Eq.~\eqref{sum}, the Hilbert space of
the $D_N$ spin Calogero model is simply the direct sum of the two Hilbert spaces associated with
two $BC_N$ models with opposite chiralities (and $b=0$).

Turning next to the scalar Hamiltonian $\Hsc$, in view of Eq.~\eqref{HscHp}
we now need to consider scalar functions of the form
\begin{equation}\label{psissc}
\psi_\bn^\ep(\bx)=\La_{\mathrm{s}}^\ep\phi_\bn(\bx)\,,\qquad\ep=\pm 1\,,
\end{equation}
where $\La_{\mathrm{s}}^\ep$ is the projector onto states symmetric with respect to permutations
and with parity $\ep$ under sign reversals. In fact, we can take as the Hilbert space of $\Hsc$
the space of symmetric functions in $L_0^2(\RR^n)$ with even parity with respect to an even number
of coordinate sign reversals. In other words, the Hilbert space of $\Hsc$ is the direct sum of its
two subspaces $\mathbf{V}_\ep^{\mathrm{s}}\equiv\La_{\mathrm{s}}^\ep L_0^2(\RR^n)$, whose elements
have parity $\ep$ under sign reversals. The functions~\eqref{psissc} form a (non-orthonormal)
basis of the corresponding subspace $\mathbf{V}_\ep^{\mathrm{s}}$ provided that either $n_i=2k_i$
for all $i$ (for $\ep=1$), or $n_i=2k_i+1$ for all $i$ (for $\ep=-1$), with $k_1\geq\cdots\geq
k_N$ in both cases. Just as before, if for each $\ep=\pm1$ we order the basis functions
$\psi_\bn^\ep(\bx)$ according to the degree $\vert\bn\vert$, the matrix of the scalar Hamiltonian
$\Hsc$ in the basis~\eqref{psissc} is expressed as a direct sum of two upper triangular matrices,
with diagonal elements $\Esc_\bn$ also given by the r.h.s.\ of~\eqref{Ens}. However, due to the
absence in this case of spin degrees of freedom, the degeneracy factor $d_\bn^\ep$ of every
quantum number $\bn$ is one. Note also that from Eq.~\eqref{Ens}, its analogue for the energies of
the scalar Hamiltonian, and the freezing trick relation~\eqref{EEE}, it follows that all the
energies of the spin chain~\eqref{cH} are integers.

Let us next compute the partition functions $\Zsc$ and $Z$ of the
models~\eqref{Hsc} and~\eqref{H}. To begin with, from now on we
shall drop the common ground state energy $E_0$ in both models, since by Eq.~\eqref{ZZZ}
it does not contribute to the partition function $\cZ$. With this convention,
the partition function of the scalar Hamiltonian $\Hsc$ is given by
\[
\Zsc(aT)=(1+q^N)\sum_{k_1\geq\cdots\geq k_N\geq 0}q^{2\vert\bk\vert}\,,
\]
where $q=\e^{-1/(k_{\mathrm B}T)}$.
The latter sum is easily recognized as the partition function
\[
\Zsc^{\mathrm{(B)}}(aT)=\prod_{i=1}^N(1-q^{2i})^{-1}
\]
of the scalar Calogero model of $BC_N$ type evaluated in Ref.~\cite{BFGR08}. We thus have
\begin{multline}
  \label{Zsc}
  \Zsc(aT)=(1+q^N)\Zsc^{\mathrm{(B)}}(aT)\\
  =(1-q^N)^{-1}\prod_{i=1}^{N-1}(1-q^{2i})^{-1}\,.
\end{multline}
We  are now ready to compute the partition function of the spin Hamiltonian $H$ in
Eq.~\eqref{H}. As for the $BC_N$ model~\cite{BFGR08}, it is convenient
to deal separately with the cases of even and odd $m$.

\subsection{Even $m$}

When $m$ is even, condition (iii) above simplifies to\smallskip

\ni (iii${}'$) \hskip2mm $s_i>0$ for all $i$.\smallskip

\ni By Eq.~\eqref{Ens}, after dropping $E_0$ the partition function of the
Hamiltonian~\eqref{H} can be written as
\begin{equation}\label{Zep}
Z(aT)=\sum_{n_1\geq\cdots\geq n_N\geq 0}d_\bn\ms q^{\vert\bn\vert}\,,
\end{equation}
where  $d_\bn$ is the spin degeneracy factor associated with the
quantum number $\bn$.
Writing
\begin{equation}
  \bn=\big(\overbrace{\vphantom{1}k_1,\dots,k_1}^{\nu_1},\dots,
  \overbrace{\vphantom{1}k_r,\dots,k_r}^{\nu_r}\big),\qquad k_1>\cdots>k_r\geq0,
  \label{neven}
\end{equation}
and using the conditions (ii) and (iii${}'$), we have
\begin{equation}\label{dn}
  d_\bn=2\prod\limits_{i=1}^r\binom{m/2}{\nu_i}\equiv 2\,d(\bnu)\,,\qquad
  \bnu=(\nu_1,\dots,\nu_r)\,,
\end{equation}
where $d(\bnu)$ is the corresponding degeneracy factor for the
$BC_N$ type of spin Calogero model~\eqref{BH} with even $m$,
and the factor of $2$ is due to the two values taken by $\ep$
in Eq.~\eqref{psis}.
Note that $\sum_{i=1}^r \nu_i=N$, so that the multi-index $\bnu$ can be
regarded as an element of the set $\cP_N$ of partitions of $N$ (taking order
into account). With the previous notation, Eq.~\eqref{Zep} becomes
\begin{align}
  Z(aT)&=2\sum_{\bnu\in\cP_N}d(\bnu)
  \sum_{k_1>\cdots>k_r\geq 0}\,q^{\sum\limits_{i=1}^r\nu_i k_i}\notag\\
  &=2Z^{\mathrm{(B)}}(aT)\,,\label{Z}
\end{align}
where
\[
Z^{\mathrm{(B)}}(aT)=q^{-N}\sum_{\bnu\in\cP_N}d(\bnu)\prod_{j=1}^r\frac{q^{N_j}}{1-q^{N_j}}\,,
\quad N_j\equiv\sum_{i=1}^j\nu_i\,,
\]
is the partition function of the $\mathrm{su}(m)$ spin Calogero model of $BC_N$ type with even
$m$, cf.~\cite{BFGR08}. {}From Eqs.~\eqref{ZZZ}, \eqref{Zsc}, \eqref{Z} and the latter expression
we finally obtain the following explicit formula for the partition function of the
$\mathrm{su}(m)$ PF chain of $D_N$ type in the case of even $m$:
\begin{equation}
  \label{cZfinal}
  \cZ(T) =
  2\prod_{i=1}^{N-1}(1-q^{2i})\sum_{\bnu\in\cP_N}d(\bnu)
  \prod_{j=1}^{\ell(\bnu)-1}\frac{q^{N_j}}{1-q^{N_j}}\,,
\end{equation}
where $\ell(\bnu)=r$ is the number of components of the multi-index $\bnu$. The latter equation
can be also written as
\begin{equation}
  \label{cZsimp}
\cZ(T) = 2\prod_{i=1}^{N-1}(1+q^{i})\sum_{\bnu\in\cP_N}d(\bnu)\,
q^{\sum\limits_{j=1}^{\ell(\bnu)-1}\kern-5pt N_j}\,
\prod_{j=1}^{N-\ell(\bnu)}(1-q^{N_j'})\,,
\end{equation}
where the positive integers $N'_j$ are defined by
\[
\big\{N_1',\dots,N'_{N-\ell(\bnu)}\big\} = \big\{1,\dots,N-1\big\}
-\big\{N_1,\dots,N_{\ell(\bnu)-1}\big\}.
\]
Note also that from the freezing trick relation~\eqref{ZZZ}, its analogous for the $BC_N$ models,
and Eqs.~\eqref{Zsc}-\eqref{Z} one easily obtains the identity
\begin{equation}\label{ZZB}
\cZ(T)=2(1+q^N)^{-1}\cZ^{(\mathrm B)}(T)\qquad\text{(even $m$)}\,,
\end{equation}
where $\cZ^{(\mathrm B)}(T)$ is the partition function of the su($m$) PF
chain~\eqref{BcH} of $BC_N$ type.

For the simplest case of spin $1/2$ chain,
we have $\nu_i=1$ for all $i$, and therefore
$\ell(\bnu)=N$, $d(\bnu)=1$ and $N_j=j$, so that Eq.~\eqref{cZsimp} simplifies
to
\begin{equation}
  \cZ(T) = 2q^{\frac12N(N-1)}\prod_{i=1}^{N-1}(1+q^i)\,,\qquad m=2\,.
  \label{Zschalf}
\end{equation}
Thus, for spin $1/2$ the spectrum of the chain~\eqref{cH} is given by
\begin{equation}
  \label{spechalf}
\cE_j = \frac12\,N(N-1)+j\,,\qquad j=0,1,\dots, \frac12\,N(N-1)\,,
\end{equation}
and the degeneracy of the energy $\cE_j$ is twice the number $Q_{N-1}(j)$ of partitions of the
integer $j$ into distinct parts no larger than $N-1$ (with $Q_{N-1}(0)\equiv 1$).

\subsection{Odd $m$}

Let us consider now the case of odd $m$. As for the $BC_N$ chain, in this case it is convenient to
slightly modify condition (i) above by first grouping the components of $\bn$ with the same parity
and then ordering separately the even and odd components. In other words, we shall write
$\bn=(\bn_{\mathrm e},\bn_{\mathrm o})$, where
\begin{align*}
  &\bn_{\mathrm e}=\big(\overbrace{\vphantom{1}2k_1,\dots,2k_1}^{\nu_1},\dots,
  \overbrace{\vphantom{1}2k_s,\dots,2k_s}^{\nu_s}\big),\\
  &\bn_{\mathrm
    o}=\big(\overbrace{\vphantom{1}2k_{s+1}+1,\dots,2k_{s+1}+1}^{\nu_{s+1}},\dots,
  \overbrace{\vphantom{1}2k_r+1,\dots,2k_r+1}^{\nu_r}\big),
\end{align*}
and
\[
k_1>\cdots>k_s\geq0,\qquad k_{s+1}>\cdots>k_r\geq0\,.
\]
The spin degeneracy factor is now
\begin{equation}\label{dnodd}
  d_\bn=d^-_s(\bnu)+d^+_s(\bnu)\equiv d_s(\bnu)\,,
\end{equation}
where $d_s^\pm(\bnu)$ is the number of independent spin states $\ket s$ satisfying conditions
(ii) and (iii) with $\ep=\pm1$, namely (cf.~\cite[Eq.~(28)]{BFGR08})
\begin{equation}
  \label{depssnu}
  d_s^\ep(\bnu)=\prod\limits_{i=1}^s\binom{\frac{m+\ep}2}{\nu_i}
  \cdot \prod\limits_{i=s+1}^r\binom{\frac{m-\ep}2}{\nu_i}\,.
\end{equation}
Calling
\[
\tilde N_j = \sum_{i=s+1}^j\nu_i\,,\qquad j=s+1,\dots,r\,,
\]
and proceeding as before, we obtain
\begin{align}
  Z(aT)&=\sum_{\bnu\in\cP_N}\sum_{s=0}^r d_s(\bnu)
  \hspace*{-10pt}\sum_{\substack{k_1>\cdots>k_s\geq 0\\k_{s+1}>\cdots>k_r\geq
      0}} \hspace*{-10pt}
  q^{\sum\limits_{i=1}^s2\nu_ik_i}q^{\sum\limits_{i=s+1}^r\nu_i(2k_i+1)}\notag\\
  &=Z^{\mathrm{(B)}}_{+}(aT)+Z^{\mathrm{(B)}}_{-}(aT)\,,\label{ZZBs}
\end{align}
where $Z^{\mathrm{(B)}}_{\pm}$ denote the partition functions of the su($m$) spin Calogero models of
$BC_N$ type~\eqref{BH} with odd $m$ and $\ep=\pm 1$. Using the expressions of $Z^{\mathrm{(B)}}_{\pm}$
derived in Ref.~\cite{BFGR08} we finally obtain
\begin{align}
Z(aT)=\sum_{\bnu\in\cP_N}\sum_{s=0}^{\ell(\bnu)}&d_s(\bnu)\,q^{-(N+N_s)}
\prod_{j=1}^s\frac{q^{2N_j}}{1-q^{2N_j}}\notag\\
&{}\times\prod_{j=s+1}^{\ell(\bnu)}\frac{q^{2\tilde N_j}}{1-q^{2\tilde N_j}}\,.
\end{align}
Substituting the previous expression and~\eqref{Zsc} into \eqref{ZZZ}, we immediately deduce the
following explicit formula for the partition function of the $\mathrm{su}(m)$ PF chain of $D_N$
type for odd $m$:
\begin{multline}
  \label{cZodd}
  \cZ(T)=(1-q^N)\prod_{i=1}^{N-1}(1-q^{2i})\sum_{\bnu\in\cP_N}
  \sum_{s=0}^{\ell(\bnu)}d_s(\bnu)\,q^{-(N+N_s)}\\
  {}\times\prod_{j=1}^s\frac{q^{2N_j}}{1-q^{2N_j}}\,\,\cdot\!\!
  \prod_{j=s+1}^{\ell(\bnu)}\frac{q^{2\tilde N_j}}{1-q^{2\tilde N_j}}\,.
\end{multline}
Equivalently (cf.~Eqs.~\eqref{Zsc} and~\eqref{ZZBs})
\begin{equation}\label{cZcZBs}
\cZ(T)=(1+q^N)^{-1}\Big(\cZ^{(\mathrm B)}_+(T)+\cZ^{(\mathrm B)}_-(T)\Big)\qquad\text{(odd $m$)}\,,
\end{equation}
where $\cZ^{(\mathrm B)}_\pm(T)$ are the partition functions of the su($m$) PF chains~\eqref{BcH} of
$BC_N$ type for odd $m$. Note that the latter formula is also valid for even $m$, since in that
case $\cZ^{(\mathrm B)}_{+}=\cZ^{(\mathrm B)}_{-}\equiv\cZ^{(\mathrm B)}$. In fact,
Eq.~\eqref{cZcZBs} can be used to verify that the expression~\eqref{cZodd} for the partition
function of the su($m$) PF spin chain of $D_N$ type is a polynomial in $q$, as should be the case
for a finite system with integer energies. To this end, recall from Ref.~\cite{BFGR08npb} that the
partition function $\cZ^{(\mathrm B)}_\ep$ can be written as
\begin{multline}
\label{cZBA}
\cZ^{(\mathrm B)}_\ep(T)=\sum_{K=0}^N q^{K\left(K-\frac12(1+\ep)\right)}\\
\times\prod_{i=K+1}^N(1+q^i)\cdot\qbinom NKq\,
\cZ_{N-K}^{\mathrm{(A)}}(q;\tfrac{m-1}2)\,,
\end{multline}
where $\cZ_{N-K}^{\mathrm{(A)}}(q;\tfrac{m-1}2)$ is the partition function of the
su$(\frac{m-1}2)$ PF spin chain of $A_N$ type with $N-K$ particles, and
\[
\qbinom NKq=\frac{(q)_N}{(q)_K(q)_{N-K}}\,,\qquad (q)_j\equiv\prod\limits_{i=1}^j(1-q^i)\,.
\]
It can be shown that both the $q$-binomial coefficient $\qbinom NKq$ and the partition function
$\cZ_{N-K}^{\mathrm{(A)}}$ are polynomials in $q$, cf.~Refs.~\cite{BUW99,Ci79}. Since all the
terms in the sum in the r.h.s.\ of Eq.~\eqref{cZBA} contain a factor of $1+q^N$ except for $K=N$, the
partition function $\cZ_\ep^{(\mathrm B)}$ can be expressed as
\[
\cZ_\ep^{(\mathrm B)}(T)=(1+q^N)\cP_\ep(q)+q^{N(N-1)}q^{\frac N2\,(1-\ep)}\,,
\]
where
\[
\cP_\ep(q)=\sum_{K=0}^{N-1} q^{K\left(K-\frac12(1+\ep)\right)}
\prod_{i=K+1}^{N-1}(1+q^i)\cdot\qbinom NKq\,
\cZ_{N-K}^{\mathrm{(A)}}
\]
is a polynomial in $q$. Inserting the latter equations into~\eqref{cZcZBs}
we immediately conclude that
\begin{equation}
\cZ(T)=\sum_{K=0}^N q^{K(K-1)}
\prod_{i=K}^{N-1}(1+q^i)\cdot\qbinom NKq\,
\cZ_{N-K}^{\mathrm{(A)}}(q;\tfrac{m-1}2)
\label{cZosimp}
\end{equation}
is a polynomial in $q$, as claimed.

\section{Statistical analysis of the spectrum}\label{stat}

In this subsection we shall take advantage of the explicit expressions for the partition function
of the su($m$) PF chain of $D_N$ type~\eqref{cH} just derived to check that its spectrum shares
the global properties of those of other spin chains of Haldane--Shastry type mentioned in the
Introduction. In practice, in order to compute the spectrum for given values of $N$ and $m$ it is
more efficient to use Eq.~\eqref{cZosimp} for odd $m$ and its analog for even $m$
\begin{equation}
  \label{cZsimpe}
  \cZ(T)=2\cZ^{(\mathrm A)}_N(q;\tfrac m2)\prod_{i=1}^{N-1}(1+q^i)\,,
\end{equation}
obtained from Eq.~\eqref{ZZB} using Eq.~(31) in Ref.~\cite{BFGR08npb}, together
with the explicit expression
\[
\cZ^{(\mathrm A)}_{K}(q;n)=\sum_{M_1+\cdots+M_{n}=K}q^{\frac12\sum\limits_{j=1}^{n}M_j(M_j-1)}
\frac{(q)_K}{(q)_{M_1}\cdots(q)_{M_{n}}}
\]
derived in Ref.~\cite{BUW99}. With the help of the previous formulas it is possible to
determine the chain's spectrum for relatively large values of $N$ and $m$; for instance, using
\mbox{\textsc{Mathematica}\texttrademark{}} on a personal computer it 
takes less than 10 seconds to evaluate the partition function in the case $N=50$ and $m=3$.

\begin{figure}[h]
\includegraphics[width=8.6cm]{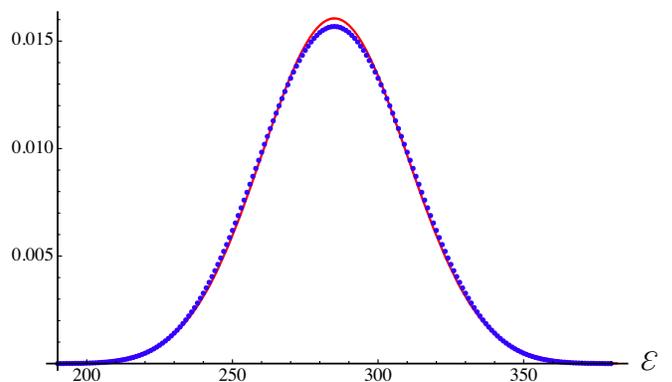}
\caption{Plot of the Gaussian distribution~\eqref{Gaussian} (continuous red line)
versus the l.h.s.\ of Eq.~\eqref{fg} (blue dots) in the case $m=2$
and  $N=20$. The root mean square error (normalized to the mean)
of the adjustment is $3.01\times 10^{-2}$.}
\label{fig:levden}
\end{figure}

In the first place, our calculations of the spectrum for a wide range of values of $m$ and $N$
show that the energies of the $D_N$ chain~\eqref{cH} form a set of consecutive integers, as is the
case for all the previously studied (non-supersymmetric) rational chains, of both $A_N$ and $BC_N$
type~\cite{Po94,BFGR08}. As to the (normalized) level density
\begin{equation}\label{f}
f(\cE)=m^{-N}\sum_{i=1}^L d_i\,\de(\cE-\cE_i)\,,
\end{equation}
where $\cE_1<\cdots<\cE_L$ are the distinct energy levels and $d_i$ is the degeneracy of $\cE_i$,
we have verified that when $N$ is sufficiently large it can be approximated with great accuracy by
the Gaussian law
\begin{equation}\label{Gaussian}
g(\cE)=\frac{1}{\sqrt{2\pi}\si}\,\e^{-\frac{(\cE-\mu)^2}{2\si^2}}
\end{equation}
with parameters $\mu$ and $\si$ given by the mean and standard deviation of the chain's spectrum.
Since the energy levels are consecutive integers, this means that
\begin{equation}\label{fg}
\frac{d_i}{m^N}\simeq g(\cE_i)\qquad (N\gg 1)\,.
\end{equation}
As an illustration, in Fig.~\ref{fig:levden} we have plotted both sides of the latter equation
in the case $m=2$ and $N=20$.

In view of the approximate relation~\eqref{fg}, it is of interest to evaluate the mean and
standard deviation of the energy in closed form for arbitrary values of $N$ and $m$. This can be
done in essentially the same way as for the $BC_N$ chain~\eqref{BcH}, using the formulas for the
traces of the spin operators $S_{ij}$, $S_i$ and $\tS_{ij}$ in Ref.~\cite{EFGR05}. Indeed, setting
\[
h_{ij}=(\xi_i-\xi_j)^{-2}\,,\qquad\tih_{ij}=(\xi_i+\xi_j)^{-2}\,,
\]
the mean energy is given by
\[
\mu=m^{-N}\tr\cH=\Big(1+\frac 1m\Big)\sum_{i\ne j}(h_{ij}+\tih_{ij})\,.
\]
The sum in the r.h.s.\ of the previous equation is clearly half the maximum energy $\cE_{\mathrm{max}}$
of the Hamiltonian~\eqref{cH}, so that by Eq.~\eqref{cEmax} we have
\begin{equation}\label{mu}
\mu=\frac12\,\Big(1+\frac 1m\Big)N(N-1)\,.
\end{equation}
Similarly, the variance of the energy is given by
\begin{multline*}
\si^2=\frac{\tr(\cH^2)}{m^N}-\mu^2=2\Big(1-\frac1{m^2}\Big)\sum_{i\ne j}(h_{ij}^2+\tih_{ij}^2)\\
-\frac4{m^2}\,(1-p)\sum_{i\ne j}h_{ij}\tih_{ij}\,,
\end{multline*}
where $p$ is the parity of $m$, and we have used Eq.~(A6) in Ref.~\cite{BFGR08}. {}From Eqs.~(A8),
(A9) and (A12) of the latter reference with $\be=0$, one easily obtains
\begin{equation}\label{si2}
\si^2=\frac1{36}\Big(1-\frac1{m^2}\Big)N(N-1)(4N+1)-\frac1{4m^2}(1-p)N(N-1).
\end{equation}

With the help of the above expressions for $\mu$ and $\si$, we can show that
Eq.~\eqref{cZcZBs} is not incompatible with the fact
that the level densities of the three chains $\cH$ and $\cH^{(\mathrm B)}$ with $\ep=\pm1$ are
approximately Gaussian for large $N$. Indeed, writing~\eqref{cZcZBs} as
\begin{equation}
\label{cZ3rel}
(1+q^N)\,\cZ(T)=\cZ^{(\mathrm B)}_+(T)+\cZ^{(\mathrm B)}_-(T)
\end{equation}
we see that the l.h.s.\ of \eqref{cZ3rel} represents the superposition of the spectrum of the
$D_N$ chain \eqref{cH} and its translation by $N$, whose level density tends to the sum of the
Gaussian $g(\cE)$ in \eqref{Gaussian} and its translate $g(\cE-N)$ as $N\to\infty$. But in this
limit we have $N\ll\si=O(N^{3/2})$, so that $g(\cE)+g(\cE+N)\simeq2g(\cE)$. Similarly, the r.h.s.\
of Eq.~\eqref{cZ3rel} is the partition function of the superposition of the spectra of the chain
Hamiltonians~\eqref{BcH} with $\ep=\pm1$, whose level density for large $N$ is approximately the sum
of two Gaussians with the same standard deviation as \eqref{Gaussian} and mean equal to $\mu+\frac
N2\big(1-\frac{\ep p}m\big)$, cf.~Ref.~\cite{BFGR08}. Hence as $N\to\infty$ the level density of
the r.h.s.\ of~\eqref{cZ3rel} is approximately given by
\[
g\Big(\cE-\frac N2\Big(1+\frac pm\Big)\Big)+g\Big(\cE-\frac N2\Big(1-\frac pm\Big)\Big)
\simeq 2g(\cE)\,,
\]
as the l.h.s.

Let us consider now the distribution of the spacings between consecutive levels in the
``unfolded'' spectrum. Recall~\cite{GMW98}, to begin with, that the unfolding of the levels
$\cE_i$ of a spectrum is the mapping $\cE_i\mapsto\eta_i\equiv\eta(\cE_i)$, where $\eta(\cE)$ is
the continuous part of the cumulative level density
\[
F(\cE)\equiv\int_{-\infty}^\cE f(\cE')\diff\,\cE'=m^{-N}\,\sum_{i;\ms\cE_i\le\cE}d_i\,.
\]
The unfolding mapping makes it possible to compare different spectra in a coherent way, since the
unfolded spectrum $\{\eta_i\}_{i=1}^L$ can be shown to be uniformly distributed regardless of the
initial level density. In our case, by the above discussion we can take $\eta(\cE)$ as the
cumulative Gaussian density~\eqref{Gaussian}, namely
\begin{equation}
\eta(\cE)=\int_{-\infty}^\cE g(\cE')\diff\,\cE'=\frac12\,
\Big[
1+\erf\Big(\frac{\cE-\mu}{\sqrt 2\si}\Big)
\Big]\,.
\end{equation}
One then defines the normalized spacings
\[
s_i=(\eta_{i+1}-\eta_i)/\De\,,\qquad i=1,\dots,L-1\,,
\]
where $\De\equiv(\eta_{L}-\eta_1)/(L-1)$ is the mean spacing of the unfolded energies,
so that $\{s_i\}_{i=1}^{L-1}$ has unit mean.
According to a well-known conjecture of Berry and Tabor, for a quantum integrable system
the density $p(s)$ of normalized spacings should be given by Poisson's law $p(s)=\e^{-s}$.
By contrast, for a system whose classical counterpart is chaotic, it is generally believed that
the spacings distribution follows instead Wigner's law $p(s)=(\pi s/2)\ms\exp(-\pi s^2/4)$,
typical of the Gaussian ensembles in random matrix theory~\cite{GMW98}.

We shall now see that the spacings distribution of the PF chain of $D_N$ type~\eqref{cH}
follows neither Poisson's nor Wigner's law, as is the case for all spin chains of HS type
studied so far~\cite{FG05,BB06,BFGR08,BFGR08epl,BFGR08npb}. More precisely, we will show that
the cumulative spacings distribution $P(s)\equiv\int_0^s p(s')\diff s'$ is
approximately given by
\begin{equation}\label{P}
P(s)\simeq 1-\frac{2}{\sqrt\pi\,\smax}\,\sqrt{\log\Big(\frac{\smax}s\Big)}\,,
\end{equation}
where $\smax$ is the maximum spacing. In fact, as proved in Ref.~\cite{BFGR08}, the previous
approximation necessarily holds for \emph{any} spectrum
$\cE_{\mathrm{min}}\equiv\cE_1<\cdots<\cE_L\equiv\cE_{\mathrm{max}}$ satisfying the following
conditions:

{\leftskip.75cm\parindent=0pt\setcounter{ex}{0}\parskip=6pt%
\cond The energies are equally spaced, \emph{i.e.}, $\cE_{i+1}-\cE_i=\de\cE$ for $i=1,\dots,L-1$.

\cond The level density (normalized to unity) is approximately given by the Gaussian
law~\eqref{Gaussian}.

\cond $\cE_{\mathrm{max}}-\mu\,,\,\mu-\cE_{\mathrm{min}}\gg\si$.

\cond $\cE_{\mathrm{min}}$ and $\cE_{\mathrm{max}}$ are approximately symmetric with respect to
$\mu$, namely
$\vert\cE_{\mathrm{min}}+\cE_{\mathrm{max}}-2\mu\vert\ll\cE_{\mathrm{max}}-\cE_{\mathrm{min}}$.

}
\noindent
Moreover, when these conditions are satisfied the maximum spacing can be estimated with great
accuracy as
\begin{equation}\label{smax}
\smax=\frac{\cE_{\mathrm{max}}-\cE_{\mathrm{min}}}{\sqrt{2\pi}\,\si}\,.
\end{equation}
It should also be noted that Eq.~\eqref{P} is valid only for spacings $s\in[s_0,\smax]$, where
\begin{equation}\label{s0}
s_0=\smax\e^{-\frac\pi4\,\smax^2}\ll\smax
\end{equation}
is the unique zero of the r.h.s.\ of~\eqref{P}  (the inequality in~\eqref{s0} follows easily
from condition (iii) and Eq.~\eqref{smax}).

We shall next check that conditions (i)--(iv) above are indeed satisfied by the spectrum of the
chain~\eqref{cH} when $N\gg1$. In fact, we already known that conditions (i) (with $\de\cE=1$) and
(ii) hold. In order to verify condition (iii), we first need to compute the maximum and minimum
energies $\cE_{\mathrm{max}}$ and $\cE_{\mathrm{min}}$. The maximum energy is clearly
\begin{equation}\label{cEmaxsum}
\cE_{\mathrm{max}}=2\sum_{i\neq j}\big[(\xi_i-\xi_j)^{-2}+(\xi_i+\xi_j)^{-2}\big]\,,
\end{equation}
whose corresponding eigenvectors are the spin states symmetric under permutations
and with parity $\pm1$ under spin reversals. Since $\cE_{\mathrm{max}}$ is independent of $m$, it is
most easily computed for the spin $1/2$ chain, whose spectrum is
explicitly given in~Eq.~\eqref{spechalf}. We thus obtain
\begin{equation}\label{cEmax}
\cE_{\mathrm{max}}=N(N-1)\,.
 \end{equation}
As to the minimum energy, Eq.~\eqref{cZ3rel} implies that
\[
\cE_{\mathrm{min}}=\min\big(\cE_{\mathrm{min},-}^{(\mathrm B)},\cE_{\mathrm{min},+}^{(\mathrm B)})\,,
\]
where the minimum energies $\cE_{\mathrm{min},\ep}^{(\mathrm B)}$ of the $BC_N$ chain~\eqref{BcH}
were computed in Ref.~\cite{BFGR08}. From Eqs.~(B1)-(B2) of the latter reference it easily follows
that $\cE_{\mathrm{min},+}^{(\mathrm B)}\le\cE_{\mathrm{min},-}^{(\mathrm B)}$, so that
\begin{multline}
  \label{cEmin}
  \cE_{\mathrm{min}}=\frac{N^2}m-\frac N2\Big(1+\frac pm\Big)+\frac1{2m}\,(m+p-2l)\\[1mm]
\times\big(l-m\ms p\,\theta(2l-m-1)\big),
\end{multline}
with
\[
l=N \mod \frac m2\,(1+p)\,.
\]
{}From Eqs.~\eqref{mu}, \eqref{si2}, \eqref{cEmax} and \eqref{cEmin}
it immediately follows that $(\cE_{\mathrm{max}}-\mu)/\si$ and $(\cE_{\mathrm{min}}-\mu)/\si$ are
both $O(N^{1/2})$ as $N\to\nobreak\infty$, so that condition (iii) is also satisfied. Finally,
from the latter equations it also follows that $\cE_{\mathrm{min}}+\cE_{\mathrm{max}}-2\mu$ is at
most $O(N)$ while $\cE_{\mathrm{max}}-\cE_{\mathrm{min}}=O(N^2)$, which proves condition (iv).

\begin{figure}[h]
\includegraphics[width=8.6cm]{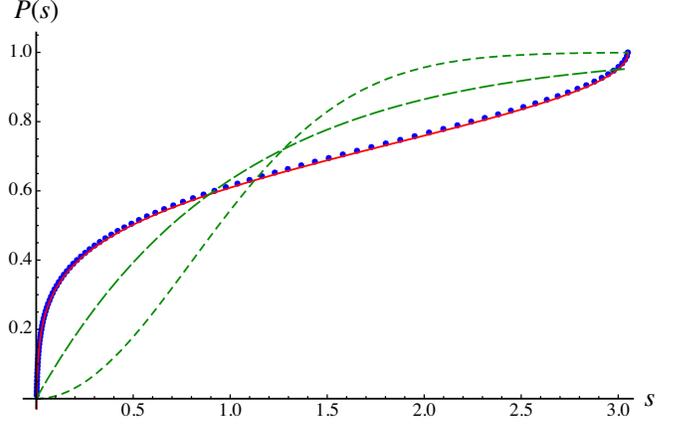}
\caption{Cumulative spacings distribution $P(s)$ and its approximation \eqref{P} (continuous red
  line) for $N=20$ and $m=2$. For convenience, we have also represented Poisson's (green, long
  dashes) and Wigner's (green, short dashes) cumulative distributions. }
\label{fig:spacings}
\end{figure}

The previous argument shows that the cumulative spacings distribution of the $D_N$
chain~\eqref{cH} should be well approximated by the r.h.s.\ of Eq.~\eqref{P} when $N$ is
sufficiently large. We have verified that~\eqref{P} is indeed in excellent agreement with the
numerical data for many different values of $N$ and $m$. For instance, in the case $N=20$ and
$m=2$ presented in Fig.~\ref{fig:spacings} the root mean square error (normalized to the mean) of
the adjustment of $P(s)$ to the r.h.s.\ of Eq.~\eqref{P} is $1.03\times 10^{-2}$, and this error
decreases to $4.69\times 10^{-4}$ when $N=100$. It should be stressed that the approximation~\eqref{P}
contains no free parameters, since the maximum spacing $\smax$ is completely determined as a
function of $N$ and $m$ by Eqs.~\eqref{si2}, \eqref{smax}, \eqref{cEmax} and~\eqref{cEmin}. In
fact, from the latter equations it immediately follows that for large $N$ the maximum spacing is
asymptotically given by
\begin{equation}
  \label{smaxasym}
  s_{\mathrm{max}}\simeq \frac3{\sqrt{2\pi}}\,\sqrt{\frac{m-1}{m+1}}\,N^{1/2}+O(N^{-1/2})\,,
\end{equation}
as for the (non-supersymmetric) PF chains of $BC_N$ type~\cite{BFGR08}.

\section{The ferromagnetic case}\label{ferro}

The ferromagnetic spin chain of $D_N$ type with Hamiltonian
\begin{equation}\label{cHF}
\cH_{\mathrm F}=\sum_{i\neq j}\bigg[
\frac{1-S_{ij}}{(\xi_i-\xi_j)^2}+\frac{1-\tS_{ij}}{(\xi_i+\xi_j)^2}\bigg]
\end{equation}
and its corresponding spin model
\[
H_{\mathrm F}=-\sum_i\pa_{x_i}^2 + \frac{a^2}4\,r^2 \,
+a\sum_{i\neq j}\bigg[
\frac{a-S_{ij}}{(x_{ij}^-)^2}+\frac{a-\tS_{ij}}{(x_{ij}^+)^2}\bigg]
\]
can be studied in much the same way as their antiferromagnetic versions~\eqref{H}-\eqref{cH}.
Since now
\begin{equation}
\label{HepFHp}
H_{\mathrm F}=H'\big|_{K_{ij}\to S_{ij},K_i\to\ep S_i}\,,
\end{equation}
we must replace the operator $\Lambda^\ep$ in Eq.~\eqref{psis} by the projector
$\Lambda^\ep_{\mathrm{s}}$ onto states \emph{symmetric} under simultaneous permutations of the
particles' spatial and spin coordinates, and with parity $\ep$ under sign reversal of coordinates
and spin. Hence condition (ii) above for the new basis states
\[
\tilde\psi_{\bn,\bs}^\ep\equiv\Lambda^\ep_{\mathrm{s}}\big(\phi_\bn(\bx)\ket\bs\big)\,,
\qquad\ep=\pm1\,,
\]  
should now read

\smallskip\noindent
(ii${}'$) $s_i\geq s_j$ whenever $n_i=n_j$ and $i<j$.

\smallskip\noindent As a result, the degeneracy factors $d(\bnu)$ and $d_s(\bnu)$ in
Eqs.~\eqref{dn} and~\eqref{dnodd} should be replaced by their ``bosonic'' versions
\[
d_{\mathrm F}(\bnu)=\prod\limits_{i=1}^r\binom{\frac m2+\nu_i-1}{\nu_i}
\]
and
\[
d_{\mathrm F,s}(\bnu)=d_{\mathrm F,s}^-(\bnu)+d_{\mathrm F,s}^+(\bnu)\,,
\]
where
\[
d_{\mathrm F,s}^\ep(\bnu)=\prod\limits_{i=1}^s\binom{\frac{m+\ep}2
+\nu_i-1}{\nu_i}
\cdot \prod\limits_{i=s+1}^r\binom{\frac{m-\ep}2+\nu_i-1}{\nu_i}\,.
\]
Therefore the partition function of the ferromagnetic $\mathrm{su}(m)$ PF
chain of $D_N$ type~\eqref{cHF} is still given by Eq.~\eqref{cZfinal} (for
even $m$) or~\eqref{cZodd} (for odd $m$), but with $d(\bnu)$ and $d_s(\bnu)$
replaced respectively by $d_{\mathrm F}(\bnu)$ and $d_{\mathrm F,s}(\bnu)$.

On the other hand, the chains~\eqref{cH} and~\eqref{cHF} are obviously related by
\begin{equation}
  \label{cHdual}
\cH_{\mathrm F}+\cH=2\sum_{i\neq
  j}\big[(\xi_i-\xi_j)^{-2}+(\xi_i+\xi_j)^{-2}\big]
=N(N-1)\,,
\end{equation}
where we have used Eqs.~\eqref{cEmaxsum}-\eqref{cEmax}. Thus the partition functions $\cZ$ and
$\cZ_{\mathrm F}$ of $\cH$ and $\cH_{\mathrm F}$ satisfy the remarkable identity
\begin{equation}
  \label{cZdual}
  \cZ_{\mathrm F}(q)=q^{N(N-1)}\cZ(q^{-1})\,.
\end{equation}

This is a manifestation of the boson-fermion duality discussed in detail in Refs.~\cite{BBHS07}
for the $\mathrm{su}(m|n)$ supersymmetric HS spin chain, since the ferromagnetic
(resp.~antiferromagnetic) chain can be regarded as purely bosonic (resp.~fermionic). For instance,
using the latter identity and Eq.~\eqref{Zschalf} we easily obtain the following expression for
the partition function of the ferromagnetic spin $1/2$ chain:
\begin{equation}\label{cZF2}
\cZ_{\mathrm{F}}(T)=2\prod_{i=1}^{N-1}(1+q^i)\,,\qquad m=2\,.
\end{equation}

With the help of the duality relation~\eqref{cZdual} and the elementary $q$-number identity
\[
(q^{-1})_K = (-1)^Kq^{-\frac12K(K+1)}\ms(q)_K
\]
it is straightforward to derive the analogs of Eqs.~\eqref{cZosimp} and \eqref{cZsimpe} for the
ferromagnetic chain~\eqref{cHF}. Calling $\cZ^{(A)}_{K,\mathrm F}(q;n)$ the partition function of
the su($n$) ferromagnetic PF chain of type A for $K$ spins, given by~\cite{BUW99}
\[
\cZ^{(A)}_{K,\mathrm F}(q;n)=\sum_{M_1+\cdots +M_n=K}\frac{(q)_K}{(q)_{M_1}\cdots (q)_{M_n}}\,,
\]
we obtain in this way
\[
\cZ(T) = 2\,\cZ^{(A)}_{N,\mathrm F}(q;\tfrac m2)\,\prod_{i=1}^{N-1}(1+q^i)
\]
for even $m$, and
\[
\cZ(T) = \sum_{K=0}^N\prod_{i=K}^{N-1}(1+q^i)\cdot\qbinom NKq\,
\cZ_{N-K,\mathrm F}^{\mathrm{(A)}}(q;\tfrac{m-1}2)
\]
for odd $m$. Finally, from the duality relation~\eqref{cHdual} it clearly follows that the
statistical properties of the spectrum of $\cH_{\mathrm F}$ are identical to those of $\cH$,
namely when $N$ is large enough the level density is approximately Gaussian, and the spacings
distribution follows Eq.~\eqref{P} with great accuracy.

\begin{acknowledgments}
This work was partially supported by Spain's DGI under grant no.~FIS2005-00752, and
by the Complutense University and Madrid's DGUI under grant no.~GR74/07-910556.
\vspace*{.5cm}
\end{acknowledgments}


\end{document}